\documentclass[12pt]{article}
\usepackage{axodraw}
\usepackage{amssymb}
\usepackage{cite}
\usepackage[]{hyperref} 
\input xy
\xyoption{all}

\begin{document}

\vspace*{0.5in}

\begin{center}

{\large\bf Bagger-Witten line bundles on moduli spaces of elliptic
curves}

\vspace*{0.2in}

Wei Gu, Eric Sharpe

Department of Physics MC 0435 \\
850 West Campus Drive\\
Virginia Tech\\
Blacksburg, VA  24061\\

{\tt weig8@vt.edu}, {\tt ersharpe@vt.edu}

$\,$

\end{center}

In this paper we discuss Bagger-Witten line bundles over moduli spaces
of SCFTs.  We review how in general they are `fractional' line bundles,
not honest line bundles, twisted on triple overlaps.
We discuss the special case of moduli spaces of elliptic curves in detail.
There, the Bagger-Witten line bundle does not exist as an ordinary line bundle,
but rather is necessariliy fractional.  As a fractional line bundle,
it is nontrivial (though torsion)
over the uncompactified moduli stack, and its restriction to the
interior, excising corners with enhanced stabilizers, is also fractional.
It becomes an honest line bundle on a moduli stack defined by a quotient
of the upper half plane by a metaplectic group, rather than 
$SL(2,{\mathbb Z})$.
We review and compare to results of recent work arguing that 
well-definedness of the worldsheet metric
implies that the Bagger-Witten line bundle admits a flat connection
(which includes torsion bundles as special cases),
and give general arguments on the existence of universal structures on
moduli spaces of SCFTs, in which superconformal deformation parameters are
promoted to nondynamical fields ranging over the SCFT moduli space.

\begin{flushleft}
June 2016
\end{flushleft}

\newpage

\tableofcontents

\newpage

\section{Introduction}

The Bagger-Witten line bundle was originally discovered as a structure
over the moduli space of scalars in four-dimensional $N=1$ supergravity
theories \cite{bw1}, and was quickly demonstrated to exist as a structure over
more general moduli spaces of two-dimensional SCFTs with $N=2$
supersymmetry \cite{ps,dist-trieste}, not only those with central
charge $c=9$ and integrally-charged chiral primaries.

Recently there have been a number of claims in the literature regarding
torsion or triviality properties of the Bagger-Witten line bundle
over either the uncompactified moduli space or the smooth part of the
uncompactified moduli space, see {\it e.g.} \cite{ks0,gkhsst}.  
The purpose of this
paper is to carefully examine these claims in the concrete
example of moduli spaces of (complex structures on)
elliptic curves, where a great deal is known
in the mathematics literature.  These moduli spaces are smooth (as stacks),
and so are very amenable to discussion. 

Briefly, we will encounter a few subtleties, but the results will
essentially agree with the predictions of \cite{gkhsst}.  
For example, in general
the Bagger-Witten line bundle, despite the name, is not actually
an honest line bundle, but rather is a `fractional' line bundle, whose
transition functions need not close on triple overlaps.  In particular,
this is true on moduli spaces of elliptic curves.  As a result, even when
we restrict to the interior of the  moduli stack, away from points with
enhanced stabilizers, it will not be possible in principle to trivialize the
Bagger-Witten line bundle.  Nevertheless, it is torsion (though
nontrivial), and the restriction
of its fourth tensor power to the interior of the moduli space is
an honest and trivializable line bundle.  

For another example,
because the chiral spectral flow operator couples to the Bagger-Witten
line bundle, it may not be possible to unambiguously define it over
the entire moduli space.  This is because fractional line bundles admit
no global meromorphic sections.  If the Bagger-Witten line bundle is
fractional, then chiral spectral flow operators can be defined in local
patches over the moduli space, but, there will be no way to globally
stitch them together over the entire moduli space
without encountering phase ambiguities.

In any event, these subtleties are consistent with the results of
\cite{gkhsst}, as we shall review, as part of a more general
discussion of universal structures over moduli spaces of SCFTs.

We begin by reviewing and extending known results on Bagger-Witten
line bundles and Fayet-Iliopoulos parameters in four-dimensional
$N=1$ supergravity theories in section~\ref{sect:4dsugrav}.
In particular, we describe how
the original Bagger-Witten story must be amended to
accomodate `fractional' line bundles, in which transition functions do not
close on triple overlaps, to encompass standard examples.  We
then discuss worldsheet realizations of these results in 
section~\ref{sect:worldsheet-basics}, and in particular
observe that the Bagger-Witten story is generic in $N=2$ SCFTs, not just
those with $c=9$ and integrally-charged chiral primaries.

In section~\ref{sect:ex:t2} we focus on the particular example of
moduli spaces of elliptic curves.  These are well-understood in the
mathematics literature, and so serve as a concrete example for discussions
of Bagger-Witten.  Over the uncompactified moduli space, the Bagger-Witten
line bundle is a fractional line bundle, not an honest line bundle,
which becomes an honest line bundle on a suitable gerbe\footnote{
Locally, a `gerbe' is an orbifold by a finite trivially-acting group.
Sigma models on gerbes have various descriptions as gauge theories and
as sigma models with restrictions on nonperturbative sectors, as we shall
review later.  In any event,
a gerbe is a special kind of stack.
} over the moduli space, one which can be presented as
\begin{displaymath}
\left[ \left( \mbox{upper half plane} \right) / Mp(2,{\mathbb Z})
\right],
\end{displaymath}
for $Mp(2,{\mathbb Z})$ the metaplectic group extending $SL(2,{\mathbb Z})$
by ${\mathbb Z}_2$.
On that gerbe over the uncompactified moduli space, it is nontrivial
(but torsion).  If one restricts to the complement of the points with
enhanced stabilizers, the fourth tensor power of the Bagger-Witten line
bundle is trivializable.  
We also make a few remarks on higher-dimensional cases and applications of
Bagger-Witten.

In section~\ref{sect:constr-Liouville} we discuss and rephrase the
recent computation of \cite{gkhsst}, which, by promoting superconformal
deformation parameters to nondynamical fields and applying supersymmetry,
argued that consistency of the worldsheet metric implies that the
Bagger-Witten line bundle must admit a flat connection (which includes
torsion bundles as special cases).  We also compare this
conclusion to results for Bagger-Witten line bundles on moduli spaces
of elliptic curves.

In section~\ref{sect:univ-overview}, we make more general remarks on
universal structures over moduli spaces of SCFTs.  We do not claim to have
a definition of a `universal SCFT,' beyond a few intuitive ideas, but
there are some general mathematical principles which can be applied,
as we review.  We also outline a proposal for how such a structure
might be built more generally in terms of a section of a sheaf of
renormalization group flows.

Finally, in section~\ref{sect:left-bundle}, we discuss other bundles
over moduli spaces of SCFTs, analogous to the Bagger-Witten line bundle,
that appear in {\it e.g.} heterotic string theories.

We will focus for most of this paper on two-dimensional SCFTs with
(2,2) supersymmetry, with the exception of
section~\ref{sect:left-bundle}.  Theories with (0,2) supersymmetry have more
complicated moduli spaces, as we shall discuss,
and so are less amenable to discussion.

\section{Review of Bagger-Witten and
Fayet-Iliopoulos}

\subsection{Four-dimensional supergravity}
\label{sect:4dsugrav}

The Bagger-Witten line bundle in four-dimensional $N=1$ supergravity
was originally introduced in \cite{bw1} in order to understand a technical
issue.  First, recall that across coordinate patches on the moduli space
$M$ (the target space of the scalars in the supergravity), the K\"ahler
potential transforms as
\begin{displaymath}
K \: \mapsto \: K \: + \: f \: + \: \overline{f} .
\end{displaymath}
In rigid supersymmetry, this is automatically a symmetry of the theory,
but in supergravity, it is not, unless combined with corresponding
actions \cite{wb}[(23.9)] on the gravitino $\psi_{\mu}$ and scalar superpartners
$\chi^i$:
\begin{displaymath}
\psi_{\mu} \: \mapsto \: \exp\left( - \frac{i}{2} {\rm Im}\, f \right) 
\psi_{\mu}, \: \: \:
\chi^i \: \mapsto \: \exp\left( + \frac{i}{2} {\rm Im}\, f \right) \chi^i .
\end{displaymath}
(Since these are chiral rotations, there are potential anomalies, as 
discussed in {\it e.g.} \cite{dist-me}.)

Now, let us consider\footnote{
We will assume that the four-dimensional
spacetime theory has exactly $N=1$ supergravity, and not more supersymmetry.
In cases with additional supersymmetry, the $R$ symmetry group would be
larger, and so across coordinate patches on the moduli space, the
$U(1)_R$ we are ultimately discussing could mix with other parts of the
$R$-symmetry group, making this story more complicated.
Instead of a Bagger-Witten line bundle, one might have a Bagger-Witten
$SU(2)_R$ bundle, for example.
} consistency across triple overlaps on the
moduli space.  
Let $\{ U_{\alpha} \}$ be a set of open patches on the moduli space $M$.
On triple overlaps $U_{\alpha} \cap U_{\beta} \cap U_{\gamma}$,
one has
\begin{eqnarray*}
f_{\alpha \beta} + \overline{f}_{\alpha \beta} \: + \:
f_{\beta \gamma} + \overline{f}_{\beta \gamma} \: + \:
f_{\gamma \alpha} + \overline{f}_{\gamma \alpha} 
& = &
(K_{\beta} - K_{\alpha}) \: + \:
(K_{\gamma} - K_{\beta}) \: + \:
(K_{\alpha} - K_{\gamma}),
\\
& = &
0.
\end{eqnarray*}
(Intuitively, this is the statement that the K\"ahler potential should
return to itself.)
This merely implies
\begin{equation} \label{eq:h-coboundary}
f_{\alpha \beta} + f_{\beta \gamma} + f_{\gamma \alpha} \: = \:
h_{\alpha \beta \gamma}
\end{equation}
for purely imaginary cochains $h_{\alpha \beta \gamma}$.  
(Since they are also holomorphic, they must be constant.)
Furthermore, on a
quadruple overlap $U_{\alpha} \cap U_{\beta} \cap U_{\gamma} \cap
U_{\delta}$,
\begin{eqnarray}
\lefteqn{
h_{\alpha \beta \gamma} - h_{\alpha \beta \delta} + h_{\alpha \gamma \delta}
- h_{\beta \gamma \delta}
} \nonumber \\
& = &
(f_{\alpha \beta} + f_{\beta \gamma} + f_{\gamma \alpha}) - 
(f_{\alpha \beta} + f_{\beta \delta} + f_{\delta \alpha}) +
(f_{\alpha \gamma} + f_{\gamma \delta} + f_{\delta \alpha}) -
(f_{\beta \gamma} + f_{\gamma \delta} + f_{\delta \beta}) ,
\nonumber \\
& = & 0,  \label{eq:h-cocycle}
\end{eqnarray}
using the convention that $f_{\alpha \beta} = - f_{\beta \alpha}$.

If there is no gauge symmetry, then we can further constraint the
$h_{\alpha \beta \gamma}$ using the fermions.  For example,
using the fact that
\begin{displaymath}
\chi^i_{\beta} \: = \: \exp\left( + \frac{i}{2} {\rm Im}\, f_{\alpha \beta}
\right) \chi_{\alpha}^i,
\end{displaymath}
we see that
\begin{eqnarray}
\chi^i_{\gamma} 
& = & \exp\left( + \frac{i}{2} {\rm Im}\, f_{\beta \gamma }
\right) \chi_{\beta}^i \: = \:
\exp\left( + \frac{i}{2} {\rm Im} \, f_{\beta \gamma }
\right)
\exp\left( + \frac{i}{2} {\rm Im}\, f_{\alpha \beta} \right)
\chi^i_{\alpha}, \nonumber \\
{\rm also} & = &
\exp\left( + \frac{i}{2} {\rm Im}\, f_{\alpha \gamma} \right)
\chi^i_{\alpha},   \label{eq:fermi-triple}
\end{eqnarray}
implying that
\begin{displaymath}
\frac{i}{2} {\rm Im}\,\left(
f_{\beta \gamma} + f_{\alpha \beta} - f_{\alpha \gamma} \right) \: = \:
\frac{i}{2} (-i) h_{\alpha \beta \gamma} \: = \:
2 \pi i n_{\alpha \beta \gamma},
\end{displaymath}
or more simply,
\begin{equation}  \label{eq:h-triple-bundle}
h_{\alpha \beta \gamma} \: = \: 4 \pi i n_{\alpha \beta \gamma},
\end{equation}
for integers $n_{\alpha \beta \gamma}$.

In this case, $\exp(- f_{\alpha \beta}/2)$ define transition functions
for an honest line bundle, and the $h_{\alpha \beta \gamma}$ define the first
Chern class, by realizing the coboundary map in the long exact sequence
associated to
\begin{displaymath}
0 \: \longrightarrow \: {\mathbb Z} \: \longrightarrow \: {\cal O} \:
\stackrel{\rm exp}{\longrightarrow} \: {\cal O}^{\times} \:
\longrightarrow \: 1.
\end{displaymath}
Furthermore, condition~(\ref{eq:h-triple-bundle}) implies that the
first Chern class must be even.

However, if there is a gauge symmetry, we have more flexibility, as 
the fermions need only close on triple overlaps up to gauge transformations,
weakening the condition~(\ref{eq:fermi-triple}) and leaving
the $h_{\alpha \beta \gamma}$ less constrained.  
(In particular, in supergravity, gauge transformations
will in general also act on the gravitino, see for example
\cite{wb}[(25.25)].)
As a result, the $h_{\alpha \beta \gamma}$ need not 
obey~(\ref{eq:h-triple-bundle}) in gauge theories, and 
so the transition functions $\exp(-f_{\alpha \beta}/2)$ need not define an
honest line bundle over the moduli space in a gauge theory,
as they need not close on triple overlaps:
\begin{displaymath}
\exp\left( - f_{\alpha \beta}/2 \right) \exp\left( - f_{\beta \gamma}/2\right)
\exp\left( - f_{\gamma \alpha}/2 \right) \: = \:
\exp\left( - h_{\alpha \beta \gamma}/2 \right) \: \neq \: 1.
\end{displaymath}
This is not merely an abstract consideration, but will arise explicitly
when studying {\it e.g.}
moduli spaces of elliptic curves, as we shall discuss in
section~\ref{sect:ex:t2}.

Let us explore this more general case, in which 
$h_{\alpha \beta \gamma} \not\in 4 \pi i {\mathbb Z}$.
Condition~(\ref{eq:h-cocycle}) means that the $\{ h_{\alpha \beta \gamma} \}$
define a 2-cocycle, and the $\{f_{\alpha \beta} \}$ define what is
known as a `twisted' bundle, encountered more commonly in disucssions of
D-branes in $B$ field backgrounds.  
Now, in general, the rank of a twisted bundle
is related to the cohomology class of the twisting cocycle
$\{ h_{\alpha \beta \gamma} \}$.  In addition, 
condition~(\ref{eq:h-coboundary}) implies 
that the $\{ h_{\alpha \beta \gamma} \}$
also define a 2-coboundary, and hence are trivial in cohomology, which is
consistent with the fact that our `twisted' bundle is rank one.
The resulting structure is sometimes called a `fractional' line bundle
(see {\it e.g.} \cite{ajmos}).

A simple example of a fractional line bundle can be constructed as follows.
Construct a projective space by quotienting ${\mathbb C}^2 - \{0\}$ by
${\mathbb C}^{\times}$ with weights $(2,2)$ rather than $(1,1)$, and 
construct a line bundle by adding another ${\mathbb C}$ factor to the
quotient, of weight $1$.  This is the line bundle
\begin{displaymath}
{\cal O}(1) \: \longrightarrow \: {\mathbb P}^1_{[2,2]},
\end{displaymath}
which is sometimes denoted ${\cal O}(1/2)$.  Its tensor square is the
ordinary line bundle ${\cal O}(1)$ on ${\mathbb P}^1$, but it itself is not an
honest line bundle.

Regardless of whether it is an honest line bundle or merely a fractional
one, we shall refer to the line bundle defined by the transition
functions $\exp\left( - f_{\alpha \beta}/2 \right)$ as the
Bagger-Witten line bundle
and denote it ${\cal L}_{\rm BW}$.

As we shall outline later (see also {\it e.g.} \cite{ajmos} and
references therein for background),
if the moduli space is replaced by a gerbe over the original moduli space,
such fractional line bundles can sometimes become\footnote{
Equivalently, a map into a gerbe is a map into the underlying space with
a restriction on allowed degrees.  If the lift of a fractional line bundle
to a gerbe is an honest line bundle on the gerbe, then a map $\phi$ into the
gerbe is a map $\phi$ into the underlying space such that $\phi^* ( h_{\alpha
\beta \gamma}) \in 2 \pi i {\mathbb Z}$, so that the pullback by $\phi$ of the 
fractional line bundle is an honest line bundle.  
See {\it e.g.} \cite{hs} for this alternative interpretation of gerbes
in terms of sigma models with restrictions on allowed nonperturbative
sectors.
} honest line bundles 
over the new gerbe moduli `space', or rather, moduli stack.
Physically, sigma models on gerbes and stacks
correspond to 
gauge theories, as discussed in {\it e.g.} \cite{gerbe,nr,msx,decomp,hs} 
for two- and
four-dimensional theories,
precisely the circumstances just outlined in which fractional line
bundles might arise in four-dimensional supergravities.
See in particular
\cite{hs} for further discussion of supergravity theories in which
the moduli `space' is a stack, and the Bagger-Witten line bundle
fractional, and the implications for fractional quantizations.

Partly as a result of their alternate interpretation as honest line bundles
on stacks, such fractional line bundles have properties that are
very similar to ordinary line bundles.  For example, given fractional line
bundles ${\cal L}$, $\widetilde{\cal L}$ defined by (logarithms of)
transition functions
$f_{\alpha \beta}$, $\tilde{f}_{\alpha \beta}$ on double overlaps and
$h_{\alpha \beta \gamma}$, $\tilde{h}_{\alpha \beta \gamma}$ on
triple overlaps, there is a tensor product ${\cal L} \otimes
\widetilde{\cal L}$ defined by (logarithms of) transition functions
\begin{displaymath}
f_{\alpha \beta} \: + \: \tilde{f}_{\alpha \beta}
\end{displaymath}
on double overlaps and
\begin{displaymath}
h_{\alpha \beta \gamma} \: + \: \tilde{h}_{\alpha \beta \gamma}
\end{displaymath}
on triple overlaps.
Thus, we can manipulate fractional line bundles for some purposes as if
they were honest line bundles.

The gravitino is in principle
a spinor-valued $C^{\infty}$ section\footnote{
Strictly speaking, the given expression is only derived for the case that
${\cal L}_{\rm BW}$ is an honest bundle.  In that case, there is a
noncanonical $C^{\infty}$ isomorphism $\overline{\cal L}_{\rm BW} \cong 
{\cal L}_{\rm BW}^{-1}$, which we have utilized above.
More generally, we should interpret the gravitino as a smooth section
of $TX \otimes \phi^* {\cal L}_{\rm BW}^{+1/2} \otimes
\phi^* \overline{\cal L}_{\rm BW}^{-1/2}$.
Similar remarks hold for the scalar superpartners.
} of $TX \otimes \phi^* {\cal L}_{\rm BW}$,
where $X$ is the four-dimensional spacetime and $\phi: X \rightarrow M$ the
scalar vevs, and the scalar superpartners are spinor-valued
sections of $\phi^*(TM \otimes {\cal L}_{\rm BW}^{-1})$.  
In this language, if the Bagger-Witten line bundle
is an honest line bundle, then the K\"ahler
form on $M$ is\footnote{
In type II compactifications, it was argued in \cite{dist-trieste} that
the K\"ahler form is a de Rham representative of the first Chern class of
the cubic tensor power, rather than the square.
} a (de Rham representative of) $c_1({\cal L}_{\rm BW}^{-2})$, 
and so from
demanding positivity of the fermion kinetic terms, 
if the Bagger-Witten line bundle ${\cal L}_{\rm BW}$
is an honest line bundle, then it
must be a negative bundle, or more generally, ${\cal }_{\rm BW}^{-1}$ must
be ample.

In this language,
the superpotential $W$ transforms as
\begin{displaymath}
W \: \mapsto \: W \exp(-f) ,
\end{displaymath} 
or equivalently
is a meromorphic section of ${\cal L}_{\rm BW}^{\otimes 2}$,
so that the Yukawa couplings 
\begin{displaymath}
\exp(K/2) \, \chi^i \chi^j D_i D_j W
\end{displaymath}
are invariant.

At this point, we should note that fractional line bundles admit neither
meromorphic sections nor higher sheaf cohomology, neither as fractional line
bundles on an underlying space (or stack), nor as fractional or honest
line bundles on a gerbe over the space or stack.  Therefore, if
${\cal L}_{\rm BW}^{\otimes 2}$ is a fractional line bundle, then as the
superpotential is a meromorphic section of ${\cal L}_{\rm BW}^{\otimes 2}$,
it does not exist globally over the moduli space.

Four-dimensional Fayet-Iliopoulos parameters\footnote{
We are referring to moduli-independent parameters.  Historically,
it was thought it was thought that they must vanish for reasons
outlined in \cite{dt}.  The paper \cite{seib10} pointed out a loophole in those
arguments, that they might be nonzero so long as they were quantized,
which was exploited and used to link Bagger-Witten and Fayet-Iliopoulos
in \cite{dist-me,hs}.  
} are closely interrelated
\cite{dist-me,hs}.
To gauge the action of a group $G$, one must specify not only how it acts
on\footnote{
In principle, we could work with either the compactified moduli space or
omit points at infinite distance.  Since $G$ acts by isometries, points at
infinite distance are merely exchanged with one another, not with points at
finite distance.
} the moduli space $M$ of scalar vevs, but also how it acts on the
Bagger-Witten line bundle.  
Briefly, under the action of the gauge group \cite{wb}[(25.14)], 
\begin{eqnarray*}
\delta \phi^i & = & \epsilon^{(a)} X^{(a) i}, \\
\delta A_{\mu}^{(a)} & = & \partial_{\mu} \epsilon^{(a)} \: + \:
f^{abc} \epsilon^{(b)} A_{\mu}^{(c)},
\end{eqnarray*}
for $X^{(a)}$ a holomorphic vector field representing the infinitesimal
group action,
the K\"ahler potential
transforms as
\begin{displaymath}
\delta K \: = \: \epsilon^{(a)} F^{(a)} \: + \:
\epsilon^{(a)} \overline{F}^{(a)},
\end{displaymath}
where $F^{(a)} = X^{(a)} K + i D^{(a)}$ \cite{wb}[(25.4)].

Thus, shifts in the imaginary part of $F^{(a)}$ correspond to Fayet-Iliopoulos
parameters.  Furthermore, fermions also transform:  the scalar superpartners,
gauginos, and gravitino transform as \cite{wb}[(25.14)]
\begin{eqnarray*}
\delta \chi^i & = & \epsilon^{(a)} \left( 
\frac{ \partial X^{(a) i} }{\partial \phi^j} \chi^j \: + \:
\frac{i}{2} {\rm Im}\, F^{(a)} \chi^i \right), \\
\delta \lambda^{(a)} & = & f^{abc} \epsilon^{(b)} \lambda^{(c)} \: - \:
\frac{i}{2} \epsilon^{(a)} {\rm Im}\, F^{(a)} \lambda^{(a)}, \\
\delta \psi_{\mu} & = & - \frac{i}{2} \epsilon^{(a)} {\rm Im}\,
F^{(a)} \psi_{\mu},
\end{eqnarray*}
and the terms proportional to ${\rm Im}\,F^{(a)}$ were interpreted in
\cite{dist-me} as describing the infinitesimal lift of the group action
to the Bagger-Witten line bundle.  (Strictly speaking, \cite{dist-me}
assumed the Bagger-Witten line bundle is an honest line bundle, but 
effectively identical considerations apply to fractional line bundles,
and result in the same counting, so we shall ignore the distinction.)

Lifts of group actions to bundles need
not exist in general and, when they do exist, are not unique.  Since
${\rm Im}\, F^{(a)} = D^{(a)}+ \cdots$, 
possible lifts correspond precisely to
choices of Fayet-Iliopoulos parameter \cite{dist-me}.
Finally, in order to lift the gauge group, not just the Lie algebra of the
gauge group, such lifts are quantized, and
in general are counted by
Hom$(G,U(1))$.

Furthermore, it was argued in
\cite{hs} that the Bagger-Witten line bundle and
Fayet-Iliopoulos parameters can be combined into a single object
over the moduli stack\footnote{
It was also observed in \cite{hs} that in principle, the four-dimensional
realization of low-energy effective sigma models on stacks can be slightly
more subtle than two-dimensional realizations.  That said, IR phenomena seem
to be usefully captured by stacks -- see {\it e.g.} \cite{js1}[appendix B]
for a discussion of four-dimensional anomalies in the language of
stacks.  All that said, if $G$ is not finite, then the stack $[M/G]$
is Artin, not Deligne-Mumford, and so formal discussions of the stack
$[M/G]$ may be even less pertinent to physics, as it is only currently known
how to physically realize sigma models on Deligne-Mumford stacks.
Nevertheless, we retain the observation above, as we think it makes
an important point.
} $[M/G]$, namely the Bagger-Witten line bundle over
the moduli stack.  This is because any vector bundle on a stack
$[M/G]$ is the same as a vector bundle on $M$ with a fixed $G$-equivariant
structure, that is, a fixed choice of $G$-action.  Thus, to define
the Bagger-Witten line bundle on $[M/G]$, we must specify precisely both
the Bagger-Witten line bundle on $M$ as well as a choice of $G$-action,
which is to say, the Fayet-Iliopoulos parameter.
Thus, Bagger-Witten and Fayet-Iliopoulos
are usefully considered to be different aspects
of a single whole.

Although the Bagger-Witten line bundle 
was originally discussed in
four-dimensional $N=1$ supergravity, it also appears much more generally
in moduli spaces of SCFTs, even in SCFTs not directly related to 
supersymmetric compactifications to four dimensions.
We shall review Bagger-Witten (and also describe Fayet-Iliopoulos)
in that more general setting next.

Before proceeding, we should also note that much of our description
above implicitly assumes that the moduli space or stack is smooth.
In fact, the moduli stack of elliptic curves that we shall discuss
in section~\ref{sect:ex:t2} is smooth.  However, in (0,2) theories
in two dimensions,
moduli spaces and stacks are often not smooth, instead exhibiting
{\it e.g.} multicritical behavior, as we shall also discuss in that
section, so (0,2) theories must be treated with more care.
For this reason, in this paper we will primarily focus on (2,2) theories 
in two dimensions.

There has also been more recent work studying global aspects of
Bagger-Witten line bundles,
see {\it e.g.} \cite{ks0,gkhsst}.  From a target-space supergravity
perspective, it is important to bear in mind in such discussions that
over the moduli space, there are loci where the four-dimensional supergravity
description breaks down.  We should also observe that that work has made
claims stemming from a need for the Ferrara-Zumino multiplet to be
well-defined; however, the transformation law for the Ferrara-Zumino
multiplet in {\it e.g.} \cite{ks1}[equ'n (2.3)] resembles that of a connection
on the Bagger-Witten line bundle, and so it may suffice for such arguments
if the Bagger-Witten line bundle admits a flat connection, rather than
necessarily be trivial.
In particular, if in some particular presentation,
the transition functions for the Bagger-Witten line
bundle happen to be constant, then the Ferrara-Zumino multiplet would be
invariant across coordinate patches, and the Bagger-Witten line bundle
would admit a flat connection.  That said, this property would not be
preserved under general bundle isomorphisms.
In any event, we will touch on these
topics in section~\ref{sect:constr-Liouville}.

\subsection{Worldsheet realizations}
\label{sect:worldsheet-basics}

Consider a moduli space of SCFTs with at least one $N=2$ symmetry algebra
in either (or both) chirality.  (This is a necessary condition for
spacetime supersymmetry.)  As one walks around loops on the moduli space
of SCFTs, the theory returns to itself up to global symmetry transformations
 -- generically, the $U(1)_R$ symmetry in the $N=2$ algebra
\cite{ps,dist-trieste}.

As a result, locally the family of SCFT's transforms under the action of
some principal $U(1)$ bundle over the SCFT moduli space, and if this is
a chiral $U(1)$, the transition functions should correspond to a 
holomorphic line bundle.
If the SCFT has any additional symmetries, that bundle may be twisted.
For simplicity, we will assume that there are no other
$R$ symmetries that the $U(1)_R$ symmetry could mix with as one moves
around the moduli space, as would happen if worldsheet supersymmetry
were enhanced beyond $N=2$.  In such cases, instead of a 
principal $U(1)_R$ bundle, one would have a larger bundle, for which the
$U(1)_R$ structure is, at best, a subbundle.

In the case that the chiral primaries in the $N=2$ algebras admit
integrally-quantized charges in conventional normalizations (another
necessary condition for spacetime supersymmetry), the
SCFTs admit a spectral flow operator.
Specifically, in the notation and
conventions of \cite{lvw}, in which the two supercharges
in the $N=2$ algebra have charge $\pm 1$ with respect to the $U(1)_R$,
the spectral flow operator ${\cal U}_{\theta}$
that rotates by $\theta$, has charge
$-(c/3)\theta$.  Thus, for example, a rotation from R to NS requires
spectral flow $\theta = 1/2$, and the corresponding operator has
charge $-c/6$.  The holomorphic top form corresponds to flow by 
$\theta=1$, which has charge $-c/3$, which (up to a sign) is the degree
of the holomorphic top form.  In any event, as a result, the
spectral flow operator ${\cal U}_{1/2}$
couples to the Bagger-Witten line bundle ${\cal L}_{\rm BW}$
over the moduli space,
and so ${\cal U}_1$, the spacetime superpotential, and
the holomorphic top form should couple to 
${\cal L}_{\rm BW}^{\otimes 2}$,
precisely matching expectations from supergravity.
(See section~\ref{sect:left-bundle} for examples of other bundles
over the SCFT moduli space.)

Ultimately the reason for the coupling of ${\cal U}_{1/2}$ to
the Bagger-Witten line bundle ${\cal L}_{\rm BW}$
is that the spectral flow operator ${\cal U}_{1/2}$
is charged under the $U(1)_R$, and also implements spacetime
supersymmetry on worldsheet vertex operators.  Thus, when the spectral
flow operator picks up a phase under the transition functions of
the $U(1)$ bundle above, the scalar superpartners and gravitinos of the
target-space supergravity also pick up phases, the same phases that
one identifies with the Bagger-Witten line bundle in target-space
supergravity.

In passing, note that the realization in SCFT is much more general
than in target-space supergravity
 -- there is no constraint on the central charge
$c$, for example, nor is there a constraint that the SCFT be interpreted
as a nonlinear sigma model.  As a result, the Bagger-Witten line
bundle ${\cal L}_{\rm BW}$ plays a much more general role in SCFT
moduli spaces than the target-space supergravity analysis of the last
section would suggest.

As a consistency check, let us consider the Yukawa couplings.
In each chiral supermultiplet, the vertex operator $V_{\chi}$ for
the fermion is obtained by applying a spectral flow operator 
${\cal U}_{1/2}$ to the
vertex operator $V_{\phi}$ for the boson, and so over the 
SCFT moduli space, $V_{\chi}$ couples to the Bagger-Witten line bundle
${\cal L}_{\rm BW}$.  For example, ${\bf \overline{27}}^3$ superpotential terms
are computed by
worldsheet three-point couplings of the form
\begin{displaymath}
\langle V_{\chi} V_{\phi} V_{\chi} \rangle
\end{displaymath}
and so we see again that $W$ is a meromorphic section of 
${\cal L}_{\rm BW}^{\otimes 2}$
over the SCFT moduli space.

The discussion above suggests, correctly, that over a moduli space
of complex structures on a Calabi-Yau threefold, 
${\cal L}_{\rm BW}^{\otimes 2}$ should be interpreted
as a (possibly fractional)
line bundle defined by holomorphic top-forms, since the spectral
flow operator ${\cal U}_1$ that couples to ${\cal L}_{\rm BW}^{\otimes 2}$ 
is also interpreted
in terms of the holomorphic top form.  We shall denote the (possibly fractional)
line
bundle\footnote{
See \cite{hodgeintro} for a very readable introduction to Hodge theory,
which encapsulates related ideas.
Definition 3.13 in that reference defines Hodge bundles of any
bidegree, related to the bundle of holomorphic top-forms above.  
The reader should note that although the entire variation of
Hodge structures carries a flat connection, the filtration consists of 
holomorphic subbundles which need not be flat \cite{tonypriv}.
} of holomorphic top-forms over the complex structure moduli space
by ${\cal L}_{\rm H}$.  
The fiber over a point of ${\cal L}_{\rm H}$ over a point of the
moduli space of complex structures of a Calabi-Yau $n$-fold
is the one-dimensional
vector space $H^{n,0}(X)$, where $X$ is the Calabi-Yau corresponding to
that point on the moduli space.  In principle, its holonomies can
be extracted from the Picard-Fuchs equation \cite{ms2a,ms2b,katzpriv}.

Thus, in particular, 
there are two
(possibly fractional) line bundles of interest over moduli spaces,
namely the Bagger-Witten line bundle ${\cal L}_{\rm BW}$ and the
line bundle of holomorphic top-forms ${\cal L}_{\rm H}$, related by
${\cal L}_{\rm H} \cong {\cal L}_{\rm BW}^{\otimes 2}$.

As an aside, the Gukov-Vafa-Witten superpotential \cite{gvw} is consistent
with the structure above.  It has the form
\begin{displaymath}
W \: = \: \int \Omega \wedge G
\end{displaymath}
where $\Omega$ is a holomorphic top-form on the Calabi-Yau, and $G$ is
a flux background.  The expression above implies that the superpotential
is a meromorphic section of the same tensor power of the Bagger-Witten
line bundle as $\Omega$ over the moduli space of complex structures,
and indeed, both are local sections of ${\cal L}_{\rm H} \cong
{\cal L}_{\rm BW}^{\otimes 2}$ over the  moduli space.

So far we have discussed worldsheet realizations of (possibly fractional)
Bagger-Witten line bundles.  In four-dimensional spacetime theories,
as previously discussed, Bagger-Witten line bundles and Fayet-Iliopoulos
parameters are closely interrelated.  However, on the worldsheet,
they necessarily become rather different.  If we have a $G$ gauge
symmetry in the spacetime theory, then that is realized on the worldsheet
as a $G$ global symmetry, and in particular, in a SCFT, as a $G$ Kac-Moody
algebra at some level.  However, if we have nonzero Fayet-Iliopoulos
parameters in the spacetime theory, then if spacetime supersymmetry is
preserved, the gauge symmetry in the target space is necessarily Higgsed,
so there is no $G$ global symmetry or Kac-Moody algebra in the worldsheet
theory, or if spacetime supersymmetry is not preserved, then we do not have a
perturbative vacuum and so we do not have a SCFT with integral $U(1)_R$
charges on the worldsheet.  In either event,
this makes Fayet-Iliopoulos parameters more difficult to understand
on the worldsheet\footnote{
There are old worldsheet-based string one-loop
computations in the literature of spacetime
D-terms, see {\it e.g.} \cite{sen1,sen2,sen3,dis,lns}.  Unfortunately it is
not completely clear to us whether those computations are describing
moduli-independent or moduli-dependent Fayet-Iliopoulos parameters, and
in principle the discussion here concerns moduli-independent cases only.
}.

\section{Examples and applications}
\label{sect:ex:t2}

In this section, we will study the Bagger-Witten line bundle
in a simple example of 
a moduli space of $(2,2)$ SCFTs, namely moduli spaces of (complex
structures on) elliptic curves.
These would not themselves result in a compactification to a four-dimensional
$N=1$ theory, but as noted above, Bagger-Witten line bundles arise
in general $N=2$ SCFTs, not just those with $c=9$ and integrally-charged
chiral primaries.  As moduli spaces of elliptic curves have been
thoroughly studied in the mathematics literature, they will provide
explicit examples we can understand in detail.
We will first review\footnote{
See for example \cite{hainrev} for more information on the mathematics
of moduli spaces of elliptic curves,
and \cite{mumford} for an older discussion of their
Picard groups.
} the mathematics, then describe the corresponding
physics.  After discussing elliptic curves, we will briefly comment
on higher-dimensional examples and applications. 

\subsection{Moduli stacks of elliptic curves}

The moduli space of elliptic curves to which we often refer in the
physics literature is the quotient of the upper half plane by 
$PSL(2,{\mathbb Z})$, and can be understood more simply as the
weighted projective stack
\begin{displaymath}
{\mathbb P}^1_{[2,3]} \: - \: {\rm point},
\end{displaymath}
where the deleted point is an ordinary non-stacky point,
corresponding to a nodal elliptic curve (morally the
large-complex-structure limit point) at $i \infty$.  Strictly speaking, the
description above replaces the quotient singularities at the corners
by stack structures, so as to get a better behaved object, but at generic
points is still a space.  This is known technically as the `coarse moduli
space,' and we shall denote it, following \cite{hainrev}, by
${\cal M}_{1,1}^{\rm red}$.  

Unfortunately, the coarse moduli space is not well-behaved for many
applications, including ours.  Consider, for example, the 
line bundle ${\cal L}_{\rm H}$ of holomorphic top-forms over the moduli space,
whose square root is the Bagger-Witten line bundle.
This line bundle is well-understood in the mathematics community,
and is known in this context as the Hodge line bundle\footnote{
More generally \cite{tonypriv}, the Hodge line bundle is the bundle whose
fibers are determinants of the holomorphic one-forms, which in the special
case of elliptic curves, happens to coincide with the line bundle of
holomorphic top-forms.
}.
Unfortunately, the Hodge line bundle does not exist over
the moduli space ${\cal M}_{1,1}^{\rm red}$ as an ordinary bundle.
Of course, locally in a small patch on the moduli space, one can often
construct the Hodge line bundle, but the basic issue is that in general,
if one covers the space (well, the orbifold) by open sets, then on triple
intersections the transition functions necessarily do
not close, an example of the
more general `fractional' structures discussed earlier in 
section~\ref{sect:4dsugrav}.   
In any event, we shall return to this matter momentarily.

Perhaps more importantly, 
it is not possible to build a `universal' elliptic curve over
the coarse moduli space, which is to say, an object fibered over the
coarse moduli space such that the fiber at any point is the elliptic curve
corresponding to that point, with the universal property that for every family
of elliptic curves parametrized by any other space $S$, there is a unique map
from $S$ to the moduli space such that the family over $S$ is identical
to the pullback of the universal curve.  This is a well-understood phenomenon 
mathematically, and ultimately is due to the fact that the `corner'
points on the coarse moduli space, the two orbifold points, describe
elliptic curves with automorphisms.  For example, we can build 
a family of elliptic curves that cannot be described in terms of a map
from $S^1$ to the coarse moduli space:  simply take the family to be
a one-dimensional family of copies of one of the elliptic curves with
automorphisms, along $S^1$, but somewhere along $S^1$, identify copies
by one of the automorphisms rather than the identity. 
See {\it e.g.} \cite{benzvirev}[section 2.3] for a readable discussion
of this issue.
In any event, we shall return to questions of existence of universal 
objects in moduli spaces of SCFT's in section~\ref{sect:univ}.

To solve these problems, mathematicians replace the coarse moduli space
by the stacky quotient of the upper half plane by $SL(2,{\mathbb Z})$
instead of $PSL(2,{\mathbb Z})$.  This quotient has a trivially-acting
${\mathbb Z}_2$ everywhere, which the stack remembers, and in fact is
a ${\mathbb Z}_2$ gerbe over the coarse moduli space
${\cal M}_{1,1}^{\rm red}$.  This new stack, the gerbe over
${\cal M}_{1,1}^{\rm red}$, is called
the fine moduli space for elliptic curves, and is denoted
${\cal M}_{1,1}$.  More explicitly, it can be described as the
weighted projective stack
\begin{displaymath}
{\mathbb P}^1_{[4,6]} \: - \: {\rm point}.
\end{displaymath}
This stack is noncompact, as it omits the point at infinity,
and has Picard group ${\mathbb Z}_{12}$.
Its compactification is denoted $\overline{\cal M}_{1,1}$, and is simply
${\mathbb P}^1_{[4,6]}$, now including the point at infinity.
The Picard group of $\overline{\cal M}_{1,1}$ is simply ${\mathbb Z}$.

The Hodge line bundle over both ${\cal M}_{1,1}$ and
$\overline{\cal M}_{1,1}$ is \cite{hainpriv}, \cite{fo}[lemma 2.5],
\cite{hainrev}[section 5.4, 6] nontrivial,
and is the generator\footnote{
In particular \cite{tonypriv}, 
the restriction of the generator of the Picard group
of $\overline{\cal M}_{1,1}$ (given by ${\mathbb Z}$) is the generator
of the Picard group of ${\cal M}_{1,1}$ (given by ${\mathbb Z}_{12}$).
Omitting the divisor at infinity quotients the Picard group by the integers,
which in this case are the subgroup generated by multiplication by 12.
The tensor square of the Hodge line bundle on ${\cal M}_{1,1}$ is the
pullback to the gerbe of the generator of the Picard group of
${\cal M}_{1,1}^{\rm red}$, which is ${\mathbb Z}_6$.
} of the Picard group in both cases.
In fact, under the ${\mathbb Z}_2$ center of $SL(2,{\mathbb Z})$,
the Hodge line bundle
has nontrivial equivariant structure. 
(As a result, 
the Hodge line bundle is
a `fractional' line bundle on the gerbe, of the sort that would
yield in principle fractional Bagger-Witten quantizations, 
as discussed in {\it e.g.} \cite{hs}.  See also \cite{ajmos} for
more information on bundles on stacks and gerbes.)
It is straightforward to see how in principle this would be the case.
Under the action of 
\begin{displaymath}
\left[ \begin{array}{cc}
a & b \\
c & d \end{array} \right] \: \in \:
SL(2,{\mathbb Z}),
\end{displaymath}
points on the complex plane are mapped as \cite{hainrev}[section 2.3]
\begin{displaymath}
z \: \mapsto \: (c \tau + d)^{-1} z,
\end{displaymath}
at the same time that the complex structure modulus is mapped to
\begin{displaymath}
\tau \: \mapsto \: \frac{a \tau + b}{c \tau + d} .
\end{displaymath}
As a result, under the center of $SL(2,{\mathbb Z})$,
$z \mapsto -z$, and so a holomorphic top-form on an elliptic curve,
which is proportional to $dz$, maps to $-dz$, and hence is odd under
the action of the ${\mathbb Z}_2$ center.

In passing, the fact that the Hodge line bundle on ${\cal M}_{1,1}$ 
is odd under the center of $SL(2,{\mathbb Z})$ is ultimately the reason
why it does not exist as a line bundle over the coarse moduli space
${\cal M}_{1,1}^{\rm red}$.
We can certainly define it locally over any one open patch on the moduli
space, but it is ambiguous up to a sign, and so when one tries to fit
together local definitions, sometimes one runs into inconsistent sign
choices which cannot be removed.  This is the same type of issue that
prevents spinors from being defined globally on non-Spin manifolds,
for example, a connection we will make more explicit shortly.

Next, let us consider the restriction of the Hodge line bundle 
${\cal L}_H$ on ${\cal M}_{1,1}$ to the
complement of the orbifold points\footnote{
We would like to thank R.~Hain for explaining the following argument to us.
}.
The moduli space has two orbifold points, the orbits of $i$ and
$\exp(i \pi/3)$ on the upper half plane.
If ${\cal L}_{\rm H}$ denotes the Hodge line bundle, then
\cite{hainpriv} sections of ${\cal L}_{\rm H}^{\otimes k}$ over the
moduli stack are 
modular forms of weight $k$.  For example, the Eisenstein series 
$G_{2m}$ for $m>1$
is a section of ${\cal L}_H^{\otimes 2m}$, reflected in
the transformation law  
\begin{displaymath}
G_{2m}\left( \frac{ a \tau + b }{ c \tau + d} \right) \: = \:
(c \tau + d)^{2m} G_{2m}(\tau),
\end{displaymath}
for
\begin{displaymath}
\left[ \begin{array}{cc}
a & b \\
c & d \end{array} \right] \: \in \: SL(2,{\mathbb Z}).
\end{displaymath}
It is easy to check that Eisenstein series are invariant under
the ${\mathbb Z}_2$ center of $SL(2,{\mathbb Z})$, corresponding to
elements $\pm I_{2 \times 2}$, hence descend to 
${\cal M}_{1,1}^{\rm red}$.
Now, $G_4$ vanishes at the orbit of 
$\exp(i \pi/3)$ and $G_6$ vanishes at the orbit of $i$, and these are
their only zeroes.  Hence, the meromorphic modular form
$G_6/G_4$ has weight two, and neither zeroes nor poles away from the two
orbifold points, hence it trivializes ${\cal L}_H^{\otimes 2}$ on the complement
of the stacky points.

The Hodge line bundle ${\cal L}_{\rm H}$ itself is slightly different.  Because
it is a `fractional'
bundle in the language of \cite{ajmos}, it admits no meromorphic sections
over ${\cal M}_{1,1}$ or 
$\overline{\cal M}_{1,1}$, not even if one restricts to the complement
of the `corner' points.  This is also reflected in the fact that there
are no nonzero modular forms of odd degree.  

We can interpret the Hodge line bundle as a spinor over the moduli space.
It can be shown that \cite{tonypriv}
\begin{displaymath}
{\cal L}_{\rm H}^{\otimes 2} \: \cong \: \Omega^1_{ {\cal M}_{1,1} }
\end{displaymath}
on the uncompactified moduli stack ${\cal M}_{1,1}$, and similarly
\begin{displaymath}
{\cal L}_{\rm H}^{\otimes 2} \: \cong \: \Omega^1_{ \overline{\cal M}_{1,1} }[
D_{\infty}]
\end{displaymath}
over the compactified moduli stack $\overline{\cal M}_{1,1}$.
Over ${\cal M}_{1,1}$, this means that ${\cal L}_{\rm H}$ is a square root
of the canonical bundle, hence a spinor on that complex one-dimensional
moduli stack, which we will see literally reflected in the
worldsheet SCFT shortly.

Now, let us turn to the corresponding physics, beginning with the role
of stacks.
Two-dimensional nonlinear sigma models with target stacks were described
in {\it e.g.} \cite{gerbe,nr,msx,decomp}, 
and four-dimensional low-energy effective
nonlinear sigma models with target stacks were discussed in
\cite{hs}.  Briefly, in ordinary sigma models,
a stack is a means of encoding a gauge theory.
The simplest examples are orbifolds:  a nonlinear sigma model on
$[X/G]$ for $G$ finite is the same as a $G$-gauged nonlinear sigma
model on $X$.  In the present case, the moduli stacks of elliptic
curves described above are locally (though not globally) orbifolds,
and usually orbifolds by trivially-acting ${\mathbb Z}_2$'s.
(Even though the ${\mathbb Z}_2$ acts trivially geometrically, both
physics and stacks still detect its presence \cite{nr}.)

Now, in the present case, since the stack structures are appearing on
the SCFT moduli space, we need to be slightly more careful, and
distinguish several possibilities.  First, if the superconformal deformation
parameters are constants, or equivalently maps from a point into the
stack, then they are indistinguishable from coordinates on an
underlying non-stacky coarse moduli space.  However, if the superconformal
deformation parameters are promoted to nondynamical fields, as we shall
see in section~\ref{sect:constr-Liouville}, 
then their behavior is more interesting,
and we see the stack structure.  For example, a map from a submanifold
$S$ of spacetime into a stack $[M/G]$ is determined by
\begin{itemize}
\item a principal $G$ bundle $E$ with connection over $S$, and
\item a $G$-equivariant map $E \rightarrow M$.
\end{itemize}
If $G$ is finite, we can recognize this as the prototype of a map
into a $G$-orbifold, as has been discussed in {\it e.g.} \cite{msx}.
A superconformal parameter on a moduli stack
that has been promoted to a nondynamical
field is then defined by data of the sort above.
(In the special case that $S$ is a point, this data specializes to a map
into an underlying coarse moduli space.)

If the superconformal deformation parameter were promoted to a dynamical
field instead of a nondynamical one, 
then we would implicitly sum over data of the form above, which would
have more interesting effects.  For example, two-dimensional sigma models
on gerbes are equivalent to sigma models on disjoint unions of
ordinary spaces \cite{decomp}.  So long as our deformation parameters
are nondynamical, we will not see that decomposition, but if they
are dynamical, then summing over the defining data above will yield
the disjoint union of spaces described in \cite{decomp}.

Physically, over the moduli space ${\cal M}_{1,1}$, the 
operator ${\cal U}_1$
is, in principle, locally a section of the Hodge line bundle
${\cal L}_{\rm H}$ of holomorphic top-forms over the moduli space.  
However, as just outlined, the Hodge line bundle admits no meromorphic
sections over the moduli space, hence the operator 
${\cal U}_1$ is
defined only locally, not globally on the moduli space.  
Specifically, it is locally a section
of a fractional line bundle, and so its transition functions only close on
triple overlaps up to higher cocycles.  

In the worldsheet SCFT,
the holomorphic top-form corresponds
to an R sector vacuum -- literally, a spinor on the moduli space, 
as we have seen mathematically.  The
failure of the Hodge line bundle to exist as an honest line bundle
over the moduli space is precisely due to a sign ambiguity in defining
the R sector vacuum across different patches on the moduli space.
Furthermore, though we have only discussed that sign ambiguity for
moduli spaces of elliptic curves, it appears to be generic in $N=2$
SCFTs, suggesting that the Hodge line bundle will typically be
fractional over moduli spaces of other $N=2$ SCFTs.

In principle
the Bagger-Witten line bundle is a square root of
the Hodge line bundle, but as the Hodge line bundle generates the
order-12 Picard group of ${\cal M}_{1,1}$, it has no square root on
${\cal M}_{1,1}$, so to
define the Bagger-Witten line bundle, we will need to replace
the moduli stack ${\cal M}_{1,1}$.

With this motivation, we can now define our next set of
stacks, which will be ${\mathbb Z}_2$ gerbes over
${\cal M}_{1,1}$, $\overline{\cal M}_{1,1}$ on which the Hodge line bundle
admits a square root.  In fact, tautologically\footnote{
We would like to thank T.~Pantev for pointing out this construction.
}, for any line bundle $L$
on any stack ${\mathfrak X}$, there is a canonically-defined 
${\mathbb Z}_2$ gerbe on ${\mathfrak X}$ such that the pullback of $L$
to this gerbe has a square root.  The class of the gerbe is the image of
$[L]$ in $H^1({\mathfrak X}, {\cal O}^{\times})$ under the 
coboundary map for the short exact sequence
\begin{displaymath}
0 \: \longrightarrow \: {\mathbb Z}_2 \: \longrightarrow \:
{\cal O}^{\times} \: \longrightarrow \: {\cal O}^{\times} \: 
\longrightarrow 1.
\end{displaymath}
We therefore define the `stringy' moduli stacks
${\cal M}_{1,1}^S$, $\overline{\cal M}_{1,1}^S$
to be the (canonically-defined) ${\mathbb Z}_2$ gerbes over
${\cal M}_{1,1}$, $\overline{\cal M}_{1,1}$, respectively, such that the
Hodge line bundle admits a square root, namely the Bagger-Witten
line bundle ${\cal L}_{\rm BW}$.  
This stack can alternatively be presented as \cite{tonypriv}
the (stacky) quotient of
the upper half plane by $Mp(2,{\mathbb Z})$, the metaplectic\footnote{
This is the unique nontrivial central extension of $SL(2,{\mathbb Z})$ by
${\mathbb Z}_2$.  Its elements can be written in the form
\begin{displaymath}
\left( 
\left[ \begin{array}{cc}
a & b \\
c & d \end{array} \right], \:
\pm \sqrt{c \tau + d } \right),
\end{displaymath}
where
\begin{displaymath}
\left[ \begin{array}{cc}
a & b \\
c & d \end{array} \right] \: \in \: SL(2,{\mathbb Z}),
\end{displaymath}
and $\sqrt{c \tau + d}$ is considered as a holomorphic function of
$\tau$ in the upper half plane.  The multiplication is defined as
\begin{displaymath}
(A, f(\cdot)) (B, g(\cdot)) \: = \: ( AB, f(B(\cdot)) g(\cdot) ).
\end{displaymath}
Furthermore, it can be shown that the metaplectic group is an extension
of $PSL(2,{\mathbb Z})$ by ${\mathbb Z}_4$, 
\begin{displaymath}
1 \: \longrightarrow \: {\mathbb Z}_4 \: \longrightarrow \:
Mp(2,{\mathbb Z}) \: \longrightarrow \: PSL(2,{\mathbb Z}) \: 
\longrightarrow \: 1,
\end{displaymath}
as the two ${\mathbb Z}_2$'s
arising in the center and extension of $SL(2,{\mathbb Z})$ do not commute
with one another, and this is the reason for the ${\mathbb Z}_4$-gerbe
structure mentioned above.
} group.
It can be shown \cite{tonypriv} that ${\cal M}_{1,1}^S$ is
a ${\mathbb Z}_4$ gerbe over ${\cal M}_{1,1}^{\rm red}$,
that the Picard group of ${\cal M}_{1,1}^S$ is
${\mathbb Z}_{24}$, and that the Bagger-Witten line bundle
${\cal L}_{\rm BW}$ generates that Picard group.
(For completeness, the Picard group of the compactification 
$\overline{\cal M}_{1,1}^S$ is ${\mathbb Z}$, and
the restriction of the generator of ${\mathbb Z}$ to the uncompactified
stack ${\cal M}_{1,1}^S$ is the Bagger-Witten line bundle.)

In more elementary language, under the action of
\begin{displaymath}
\left[ \begin{array}{cc}
a & b \\
c & d \end{array} \right] \: \in \: SL(2,{\mathbb Z}),
\end{displaymath}
the holomorphic top-form on an elliptic curve transforms as
\begin{displaymath}
dz \: \mapsto \: (c \tau + d)^{-1} \, dz,
\end{displaymath}
hence the spectral flow operator ${\cal U}_{1/2}$, which transforms
as $\sqrt{dz}$, must transform under $SL(2,{\mathbb Z})$ as
\begin{displaymath}
{\cal U}_{1/2} \: \mapsto \: \pm \frac{1}{\sqrt{c \tau + d}} \, {\cal U}_{1/2}.
\end{displaymath}
As a result, the spectral flow operator ${\cal U}_{1/2}$ is
well-defined only over the stacky quotient of the upper half plane
by $Mp(2,{\mathbb Z})$, which is ${\cal M}^S_{1,1}$.

As another consistency check, one can show that ${\cal L}_{\rm BW}^{-1}$
is ample, which (as mentioned earlier) is necessary for 
positivity of the
fermion kinetic terms in general.
We can understand this as follows.
For (Deligne-Mumford) stacks \cite{tonypriv}, a line bundle is ample
if and only if some power is a pullback of an ample bundle from the
underlying coarse moduli stack.  In the present case, the generator of
the Picard group on the coarse moduli space ${\cal M}_{1,1}^{\rm red}$
is ample, hence ${\cal L}_{\rm BW}^{-1}$ on ${\cal M}_{1,1}^S$ is ample,
as its 
fourth tensor power is the pullback of an ample line bundle on
${\cal M}_{1,1}^{\rm red}$.

We conjecture that the stacks ${\cal M}^S_{1,1}$,
$\overline{\cal M}^S_{1,1}$
are fine moduli stacks (for the complex structure moduli)
for the superconformal field theory
(thus the $S$ superscript).  Of course, we do not know how to define
a `universal SCFT' over such a moduli stack (though see 
section~\ref{sect:univ} for some general observations.)
In any event, let us at least extend this conjecture to include K\"ahler
moduli, in a way consistent with T-duality.

As is well-known, the moduli space of $N=2$ SCFT's describing strings
on elliptic curves can be described approximately as
$O(2,2;{\mathbb Z}) \verb'\' O(2,2) / ( O(2) \times O(2) )$.
Briefly, the $O(2,2)$ decomposes, and as a result this can be written
as, modulo ${\mathbb Z}_2$'s, a product of two copies of
$PSL(2,{\mathbb Z}) \verb'\' SL(2) / O(2)$, which can be identified\footnote{
One maps $SL(2)/U(1)$ into the upper half plane
by mapping
\begin{displaymath}
\left[ \begin{array}{cc}
a & b \\ c & d \end{array} \right] \: \in \: SL(2,{\mathbb R}) \: \mapsto
\:
\frac{a i + b}{c i + d}.
\end{displaymath}
Note that
\begin{displaymath}
{\rm Im}\, \frac{ai+b}{ci+d} \: = \: \frac{ad-bc}{c^2 + d^2} \: = \:
\frac{1}{c^2+d^2} \: > \: 0,
\end{displaymath}
so the image lies in the upper half plane.
Moreover, the $U(1)$ coset acts as
\begin{displaymath}
\left[ \begin{array}{cc}
a & b \\ c & d \end{array} \right] 
\cdot \left[ \begin{array}{cc}
\cos \theta & \sin \theta \\
- \sin \theta & \cos \theta \end{array} \right] \: = \:
\left[ \begin{array}{cc}
a \cos \theta - b \sin \theta & a \sin \theta + b \cos \theta \\
c \cos \theta - d \sin \theta & c \sin \theta + d \cos \theta
\end{array} \right],
\end{displaymath}
and so its image under the map above is
\begin{displaymath}
\frac{a e^{-i \theta} i + b e^{-i \theta}
}{
c e^{-i \theta} i + d e^{-i \theta}
} \: = \:
\frac{a i + b}{c i + d},
\end{displaymath}
hence the map descends to $SL(2,{\mathbb R})/U(1)$, and in fact, it is
easy to see that it also descends to $SL(2,{\mathbb R})/O(2)$.
Furthermore, it is straightforward to check that
ths map defines an isomorphism
\begin{displaymath}
SL(2,{\mathbb R})/O(2) \: \cong \: PSL(2,{\mathbb R})/U(1) \: \cong \:
\mbox{upper half plane}.
\end{displaymath}
See \cite{psa-modt2} for more information.
}
 with
two copies of the coarse moduli space of an elliptic curve.
One copy corresponds to the complex structure moduli, the other to the
K\"ahler moduli.

If we let $\tau$ denote the complex structure parameter and 
$\sigma$ the K\"ahler structure parameter, then being more careful about
finite group quotients, the coarse moduli space
of SCFT's is of the form \cite{psa-modt2,dvv,gmr} 
\begin{displaymath}
\left( {\cal M}_{1,1}^{\rm red} \times {\cal M}_{1,1}^{\rm red} \right) /
{\mathbb Z}_2 \times {\mathbb Z}_2
\end{displaymath}
The resulting moduli space contains quotient singularities along
complex codimension one subvarieties, corresponding to where either
$\sigma$ or $\tau$ is $i$ or $\exp(\pi i/3)$, as well as two
points of maximal enhanced symmetry ($SU(3)$, $SU(2) \times SU(2)$),
where
$(\sigma, \tau) = (\exp(\pi i /3), \exp(\pi i/3))$, $(i,i)$,
lying at the intersection of the singular subvarieties.
In this spirit, it is natural to conjecture that the complete
fine moduli space
of $N=2$ SCFTs on elliptic curves is of a similar form,
a quotient of a product of two copies of ${\cal M}_{1,1}^S$,
one copy for complex moduli, another for K\"ahler moduli.
However, as discussed in section~\ref{sect:univ}, at the moment
we do not have a precise definition of a `universal
SCFT,' so we leave this conjecture for future work.

\subsection{Outline of higher dimensions}

In higher dimensions, less is known in general.
For example,
fine moduli spaces of complex structures may or may not be gerbes over
coarse moduli spaces -- it is not clear whether
the gerbe structure we have seen for elliptic
curves will generalize.  For an abelian variety of dimension $n$
({\it i.e.} a torus $T^{2n}$) \cite{tonypriv}, 
the ${\mathbb Z}_2$ that acted nontrivially
on holomorphic one-forms on elliptic curves, acts by $(-)^n$ on
holomorphic $n$-forms.  One might expect that the fine moduli space
of complex structures may have a ${\mathbb Z}_2$ gerbe structure,
but for $n$ even, the line bundle of holomorphic top-forms may be an
honest bundle, whereas for $n$ odd, it may be fractional.

That said, one particularly pertinent reference is
\cite{eva-ed}.  This paper discussed old work \cite{dsww1,dsww2}
in which it had been argued that (0,2) worldsheet
moduli were generically lifted by worldsheet instanton corrections,
via a study of a single worldsheet instanton.
Their computations could not be translated to topological field theory
computations, and so were difficult to analyze directly.
The paper \cite{eva-ed} argued that in many cases, multiple worldsheet
instanton contributions will cancel out, leaving the (0,2) moduli.
There has since been a significant number of papers on the subject,
see {\it e.g.} \cite{berglundetal,beasw,bertpless,bkos,bkos2,bkos3,ap-het} 
and references therein for a sample.

In any event, the paper \cite{eva-ed} studied worldsheet instanton corrections
in heterotic strings by working on the target space, and using the fact that
the four-dimensional spacetime $N=1$ superpotential is a section\footnote{
In passing, there is a technical subtlety here which was unknown at the
time \cite{eva-ed} was written, namely that if ${\cal L}_{\rm H}$ is
a fractional line bundle, then the superpotential cannot exist globally
as ${\cal L}_{\rm H}$ admits no meromorphic sections.  At best, one could
only write down superpotentials in local patches on the moduli space.
Globally, there would be no unambiguous way to assign a superpotential
to every point on the moduli space -- one would have different choices
related by phases.
} of 
${\cal L}_{\rm H} \cong {\cal L}_{\rm BW}^{\otimes 2}$ 
over the compactified SCFT moduli space.

In essence, the authors of \cite{eva-ed}
argued that the spacetime superpotential for neutral moduli
should vanish, and hence those moduli are unobstructed.  Interestingly,
as a consistency check they also studied the superpotential for charged
matter couplings, which do not vanish and in fact encode, on the (2,2)
locus, Gromov-Witten invariants.  Those contributions diverge at singular
loci on the moduli space, reflecting divergences in the worldsheet instanton
sums, which in the present context are interpreted as reflecting the
fact that ${\cal L}_{\rm BW}^{\otimes 2}$
is a negative line bundle, hence any nonzero (meromorphic) section must
have poles.  On the (2,2) locus, the restriction of the Bagger-Witten
line bundle to the complement of those singular loci might be trivial,
but over the compactified moduli space, its structure is tied by the
arguments of \cite{eva-ed} to Gromov-Witten theory.

It is tempting to use expressions for holomorphic top-forms to compute
the Bagger-Witten line bundle and the line bundle of holomorphic
top-forms over compactified moduli spaces,
but this does not quite work.  For completeness, let us describe both the
idea and how this fails.
First, let us explain what we mean by expressions for holomorphic 
top-forms.
For example, for elliptic curves constructed as degree-three polynomials
$p$ in ${\mathbb P}^2$, in a patch with affine coordinates
$x$, $y$ on ${\mathbb P}^2$, the holomorphic top-form $\Omega$ is
\begin{equation}  \label{eq:hol-topform-curve}
\Omega \: \propto \: \frac{dx}{\partial p/\partial y}.
\end{equation}
Now, the coefficients of $p$ act as homogeneous coordinates on the (coarse)
moduli space of complex structures, which can be constructed by quotienting
symmetries.  It is tempting to argue that since $\Omega$ is of negative
degree in those coefficients, it must therefore couple to a (fractional) line
bundle, of negative degree if it is a line bundle, which would be consistent
with earlier observations.  Unfortunately\footnote{
We would like to thank D.~Morrison for explaining this matter to us.
}, the 
expression above could be multiplied by a meromorphic section of a different
line bundle over the compactified
moduli space, one with zeroes and poles at singular and limiting points,
resulting in an $\Omega$ that is equivalent over the smooth part of the moduli
space but which couples to a different (fractional) bundle over the
compactified moduli space.  Put another way, to uniquely specify the
line bundle of holomorphic top-forms
over the moduli space requires a specification of its
behavior near limiting points, and expressions such 
as equation~(\ref{eq:hol-topform-curve}) make implicit assumptions and so
do not provide a unique answer.

In passing, we should also observe that existence of 
finite nonzero volumes of
CFT moduli spaces computed in {\it e.g.} \cite{bckmp,moore-vol} 
would appear to suggest that 
over compactified moduli spaces, Bagger-Witten line bundles may be
nontrivial.

\section{Constraints implied by worldsheet metric}
\label{sect:constr-Liouville}

Recently the paper \cite{gkhsst} used worldsheet arguments to constrain
the Bagger-Witten line bundle.
They used a variation of superconformal perturbation theory,
in which the deformation parameters are promoted to nondynamical fields,
and derived a constraint on the Bagger-Witten line bundle from the
observation that the worldsheet Liouville field has no symmetry transformations
and should be invariant across open patches on the moduli space.
In this section we will carefully work through those arguments and see
how they imply that the (possibly fractional)
Bagger-Witten line bundle should admit a flat connection.

For any fixed initial SCFT, the
paper \cite{gkhsst} performed superconformal perturbation theory,
but with the parameters promoted to non-dynamical
functions on the worldsheet.  This enabled
them to analyze perturbation theory as more nearly a low-energy
effective field theory.  Since the fields are non-dynamical,
they can be restricted to any open patch on the SCFT moduli space.
(See also {\it e.g.} \cite{fk0,fk1} for other references that also
promote the superconformal deformation parameters to non-dynamical fields.)

For simplicity, for the moment we shall assume (2,2) worldsheet
supersymmetry.  (We shall comment on subtleties in the more
general (0,2) case later.)
In a (2,2) supersymmetric theory, it is most natural for one of the complex
and K\"ahler moduli to be described by ordinary chiral multiplets, and the
other to be described by twisted chiral multiplets.  The result is a 
moduli space with a generalized complex structure in Hitchin's sense.
An example of the resulting structure is implicit in the
`ur-theory' described implicitly in \cite{daveronenmir}.

There is a universal contribution involving a supersymmetric multiplet
denoted $\Sigma$.  
In superconformal gauge, its bosonic part has the form
\cite{gkhsst}[appendix C.1]
\begin{displaymath}
\Sigma \: = \: \sigma + i a \: + \: \cdots,
\end{displaymath}
where $\sigma$ is the conformal factor in the worldsheet metric and $a$ is
defined by a non-dynamical $U(1)$ gauge field $A_{\mu}$ over the
moduli space (coupling to the Bagger-Witten line bundle), as
\begin{displaymath}
A_{\mu} \: = \: \epsilon_{\mu \nu} \partial^{\nu} a.
\end{displaymath}
(This definition of $a$ implicitly forces constraints on the $U(1)_R$
bundle over the moduli space, as we shall discuss momentarily.)
They derive that the correct kinetic term for $\Sigma$, which
encodes the supersymmetrization of the conformal anomaly, has the form
(in superconformal gauge) \cite{gkhsst}[equ'n (3.11)]
\begin{equation}  \label{eq:univ-action}
\int d^2x \int d^4 \theta \left( \Sigma + \overline{\Sigma} - \frac{6}{c} K
\right)^2
\end{equation}
where $K$ is the K\"ahler potential on (that part of) the moduli space.

The action~(\ref{eq:univ-action}) above encodes a kinetic term for $\Sigma$,
namely
\begin{displaymath}
\int d^4 \theta \, \overline{\Sigma} \Sigma ,
\end{displaymath}
as well as counterterms that arise when transforming either $\Sigma$ or $K$,
such as
\begin{displaymath}
\int d^4 \theta \, \overline{\Sigma} F .
\end{displaymath}
In particular, since $\overline{D}^2 \overline{\Sigma} = {\cal R}$, 
the chiral curvature superfield, the line above
equals \cite{gkhsst}[equ'n (3.3)]
\begin{displaymath}
\int d^2 \theta \, {\cal R} F ,
\end{displaymath}
which was used in \cite{gkhsst} to motivate the action~(\ref{eq:univ-action}).

Manifestly, the action~(\ref{eq:univ-action}) is formally invariant under
\begin{displaymath}
K \: \mapsto \: K + F + \overline{F}, \: \: \:
\Sigma \: \mapsto \: \Sigma + \frac{6}{c} F,
\end{displaymath}
for any chiral superfield $F$.
However, this would also modify the worldsheet metric, as it would shift
$\sigma$, and the lack of gauge transformations of $\sigma$ plays
a crucial role.

Let us study more closely the allowed transformations of $\Sigma$.
Consider a gauge transformation of the background $U(1)$ gauge field
(as would {\it a priori} happen across open patches on the moduli space)
\begin{displaymath}
A_{\mu} \: \mapsto A_{\mu} \: + \: \partial_{\mu} \Lambda.
\end{displaymath}
Define $\tilde{\Lambda}$ by,
\begin{displaymath}
\partial_{\mu} \Lambda \: = \: \epsilon_{\mu \nu} \partial^{\nu} 
\tilde{\Lambda},
\end{displaymath}
or in components,
\begin{eqnarray*}
\partial_1 \Lambda & = & \partial^2 \tilde{\Lambda}, \\
\partial_2 \Lambda & = & - \partial^1 \tilde{\Lambda},
\end{eqnarray*}
which the reader will recognize as being more-or-less the
Cauchy-Riemann equations.
Note that existence of $\tilde{\Lambda}$ implies that
\begin{displaymath}
\partial_{\mu} \partial^{\mu} \Lambda \: = \: 0,
\end{displaymath}
hence
\begin{displaymath}
\Lambda \: = \: f(z) \: + \: f(z)^*,
\end{displaymath}
(using both the fact that it is real and it solves Laplace's equation,)
so the field $a$ is only defined globally (as a section of an affine bundle)
if
the original $U(1)$ bundle over the moduli space, to which $A_{\mu}$ couples,
is determined by a holomorphic bundle.  As the Bagger-Witten line bundle
is indeed a holomorphic line bundle over the moduli space, this constraint
will always be satisfied.
Then, for the $\Lambda$ above,
\begin{displaymath}
\tilde{\Lambda} \: = \: {\rm Im}\left( f(z) \: - \: 
f(z)^* \right)
\end{displaymath}
and under the gauge transformation above,
$a \mapsto a + \tilde{\Lambda}$.
In particular, if $A_{\mu}$ coupled to a $U(1)_V$ transformation,
then $a$ undergoes a $U(1)_A$ transformation, and vice-versa.

Now that we have examined the allowed gauge transformations, let us
describe how these fit together systematically.  For the moment,
we will only work on open covers of the moduli space, as SCFT perturbation
theory can certainly describe open covers; in section~\ref{sect:univ},
we will discuss potential obstructions to building a `universal SCFT,'
with parameters ranging over the entire moduli space.  Let $U_{\alpha}$ be
an open cover of the smooth part of the moduli space.
(For reasons we shall discuss later, we exclude singular points from the
discussion.)  Let $\Sigma_{\alpha}$ denote the $\Sigma$ field
over $U_{\alpha}$, and $K_{\alpha}$ the K\"ahler potential across
$U_{\alpha}$.  

We can glue these fields together across coordinate patches via
field redefinitions.
Since $\sigma \propto \Sigma + \overline{\Sigma}$ 
does not admit transformations across coordinate patches,
we need to find a set of field redefinitions
\begin{displaymath}
\Sigma_{\alpha} \: \mapsto \: \Sigma_{\alpha} + C_{\alpha},
\end{displaymath}
for superfields $C_{\alpha}$ on each patch, such that on every
overlap $U_{\alpha} \cap U_{\beta}$ on the moduli space,
\begin{equation}  \label{eq:const-K}
\left.
\left( \Sigma_{\alpha} + \overline{\Sigma}_{\alpha}
+ C_{\alpha} + \overline{C}_{\alpha}\right) \right|_{U_{\alpha} \cap U_{\beta} }
\: = \:
\left. \left( \Sigma_{\beta} + \overline{\Sigma}_{\beta} + C_{\beta} 
+ \overline{C}_{\beta} \right) \right|_{U_{\alpha}
\cap U_{\beta} }.
\end{equation}
The Liouville field defined by 
\begin{equation}  \label{eq:sum-patch}
\Sigma_{\alpha} + \overline{\Sigma}_{\alpha} + C_{\alpha} 
+ \overline{C}_{\alpha}
\end{equation}
on each
patch on the moduli space will then be globally
well-defined, and we can then
talk about a globally-defined ``$\Sigma + \overline{\Sigma}$'' whose
restriction to open patch $U_{\alpha}$ is the sum~(\ref{eq:sum-patch}) above.

As previously remarked, on an overlap
$U_{\alpha} \cap U_{\beta}$,
the action~(\ref{eq:univ-action}) is invariant
under
\begin{displaymath}
K_{\alpha} \: \mapsto \: K_{\alpha} 
+ F_{\alpha \beta} + \overline{F}_{\alpha \beta}, \: \: \:
\Sigma_{\alpha} \: \mapsto \: \Sigma_{\alpha} + \frac{6}{c} F_{\alpha \beta},
\end{displaymath}
where $F_{\alpha \beta}$ is a non-dynamical chiral superfield on
$U_{\alpha} \cap U_{\beta}$ on the moduli space whose
bosonic components are the
(logarithms of the)
transition functions $f_{\alpha \beta}$ of the Bagger-Witten line bundle.

Suppose, for example, that 
\begin{displaymath}
F_{\alpha \beta} \: = \: \left. C_{\alpha} \right|_{U_{\alpha} \cap U_{\beta}}
\: - \: 
\left. C_{\beta} \right|_{U_{\alpha} \cap U_{\beta} },
\end{displaymath}
then $F_{\alpha \beta}$ can be absorbed into field redefinitions
\begin{displaymath}
\Sigma_{\alpha} \: \mapsto \Sigma'_{\alpha} \: \equiv \:
\Sigma_{\alpha} \: + \: \frac{6}{c}
C_{\alpha}, \: \: \:
\Sigma_{\beta} \: \mapsto \: \Sigma'_{\beta} \: \equiv \:
\Sigma_{\beta} \: + \: \frac{6}{c}
C_{\beta},
\end{displaymath}
so that after field redefinitions, the $\Sigma'$s are invariant across
overlaps:
\begin{displaymath}
\left. \Sigma'_{\alpha} \right|_{U_{\alpha} \cap U_{\beta}}
\: = \: \left. \left( \Sigma_{\alpha} + \frac{6}{c} C_{\alpha} 
\right) \right|_{U_{\alpha} \cap U_{\beta} }
\: \mapsto \: 
\left. \Sigma'_{\beta} \right|_{U_{\alpha} \cap U_{\beta}} \: = \:
\left. \left( \Sigma_{\beta} + \frac{6}{c} C_{\beta} \right)
\right|_{U_{\alpha} \cap U_{\beta} }.
\end{displaymath}
In this case, the Bagger-Witten line bundle would be holomorphically
trivial.

However, a weaker condition will suffice to leave the 
action~(\ref{eq:univ-action}) 
invariant.  Suppose instead that
\begin{displaymath}
F_{\alpha \beta} \: = \: \tilde{F}_{\alpha \beta} \: + \:
\left. C_{\alpha} \right|_{U_{\alpha} \cap U_{\beta}} \: - \:
\left. C_{\beta} \right|_{U_{\alpha} \cap U_{\beta} },
\end{displaymath}
where $\tilde{F}_{\alpha \beta}$ is pure imaginary (and hence constant).
In this case, the $C_{\alpha}$'s can be absorbed into field
redefinitions of the $\Sigma_{\alpha}$'s, as
\begin{displaymath}
\Sigma_{\alpha} \: \mapsto \Sigma'_{\alpha} \: \equiv \:
\Sigma_{\alpha} \: + \: \frac{6}{c} 
C_{\alpha}, \: \: \:
\Sigma_{\beta} \: \mapsto \: \Sigma'_{\beta} \: \equiv \:
\Sigma_{\beta} \: + \: \frac{6}{c}
C_{\beta},
\end{displaymath}
so that on overlaps,
\begin{displaymath}
\left. \Sigma'_{\alpha} \right|_{ U_{\alpha} \cap U_{\beta} }
 \: \mapsto \:
\tilde{F}_{\alpha \beta} \: + \: \left. \Sigma'_{\beta}
\right|_{ U_{\alpha} \cap U_{\beta} }.
\end{displaymath}
Although $\Sigma'$ is no longer invariant across overlaps,
the sum $\Sigma' + \overline{\Sigma'}$ is invariant, 
satisfying the condition~(\ref{eq:const-K})
and so the Liouville
field is still well-defined globally.
This is clearly the general case consistent with making the Liouville
field well-defined globally, and this only requires that the
Bagger-Witten line bundle admit a flat connection, but not necessarily be
trivial.

At this point, let us make two observations.
\begin{itemize}
\item First, the Bagger-Witten line bundle
${\cal L}_{\rm BW}$, with transition functions
$\exp\left( - f_{\alpha \beta}/2\right)$, might not be an honest line
bundle.  We have described a constraint on the $f_{\alpha \beta}$ on
the double overlaps $U_{\alpha \beta}$, but we have not given
a constraint on the cocycles $h_{\alpha
\beta \gamma}$ on triple overlaps beyond being
pure imaginary, the only constraint derived earlier.
If those cocycles $h_{\alpha \beta \gamma} \not\in
2 \pi i {\mathbb Z}$, then ${\cal L}_{\rm BW}$ will be fractional.
\item Over the moduli space, $\exp\left( \Sigma_{\alpha} \right)$
transforms as a local section of ${\cal L}_{\rm BW}^{- 6/c}$.
In general, depending upon ${\cal L}_{\rm BW}$ and the ratio
$6/c$, this may be a fractional line bundle.  If that should be the
case, then $\exp\left(\Sigma_{\alpha}\right)$ can only be defined 
locally on the moduli space -- the worldsheet metric will be well-defined,
but the imaginary part of $\Sigma$ will only be defined in patches.
\end{itemize}

So far we have implicitly assumed that the moduli space over which we
are working is smooth, or rather that we are working on the complenent
of singular points or points at infinite distance.  (In particular,
in SCFT's obtained from Calabi-Yau's, there will more or less always
be singularities and (large-radius) points at infinite distance.)
However, formally we would think the argument above should also apply to smooth
Deligne-Mumford stacks, following the prescription in \cite{hs},
as such stacks can be covered by smooth open patches on which one
can perform differential geometry in the usual fashion.

Let us compare the claim above to results for moduli spaces of
elliptic curves, discussed in section~\ref{sect:ex:t2}.
There, we saw that over the uncompactified moduli space,
the Bagger-Witten line bundle is nontrivial (and fractional),
but torsion, satisfying the constraint above.  If we excise the two
points of enhanced stabilizer from the uncompactified moduli space,
the situation is similar:  the restriction of
${\cal L}_{\rm BW}^{\otimes 4} \cong {\cal L}_{\rm H}^{\otimes 2}$
is trivializable, as discussed in section~\ref{sect:ex:t2},
but as the Bagger-Witten line bundle ${\cal L}_{\rm BW}$ is not
an honest line bundle, we cannot define a section, not even over this
excised region, and so it is not quite correct to say that its restriction
is trivial.

In passing, in recent discussions of supersymmetric localization,
it has been argued that the $S^2$ partition function
$Z \propto \exp(-K)$.  Clearly, this transforms over the moduli
space as a $C^{\infty}$ section of 
\begin{displaymath}
{\cal L}_{\rm H} \otimes \overline{\cal L}_{\rm H} \: = \:
{\cal L}_{\rm BW}^{\otimes 2} \otimes 
\overline{\cal L}_{\rm BW}^{\otimes 2}.
\end{displaymath}
However, under the circumstances we have discussed here,
the transition functions of ${\cal L}_{\rm BW}$ can be chosen to be pure
imaginary, so that $\overline{\cal L}_{\rm BW} \cong {\cal L}_{\rm BW}^{-1}$,
and so the line bundles above cancel out, making $\exp(-K)$ a function over
the moduli space (after suitable redefinitions on local patches).

For more or less the same reasons, the left-right symmetric spectral
flow operator should be unambiguously defined.  This is pleasant\footnote{
We would like to thank I.~Melnikov for making this observation.
} as the
operator shows up in diagonal modular invariants of (2,2) SCFTs, and
corresponds to the unique NS-NS ground state.

So far we have focused on (2,2) theories.  In (0,2) theories, there is
an additional subtlety, namely the existence of multicritical points,
{\it i.e.}
subvarieties connecting sections of the moduli space of different 
dimension, joined at well-behaved SCFT's.  This cannot happen in (2,2)
theories, as it would require the chiral ring to jump, and the chiral
ring is well-known to have the same dimension across the (2,2) SCFT
moduli space. 
However, in (0,2) theories, it is much more common.  For example,
if we take the (2,2) quintic and add suitable left-moving fields,
the resulting (0,2) theory gains a $SO(10)$ branch.

\section{Universal structures over moduli spaces of SCFTs}
\label{sect:univ-overview}

\subsection{Potential obstructions to existence of universal SCFTs}
\label{sect:univ}

It would be natural to try to interpret the results of 
section~\ref{sect:constr-Liouville} 
in terms of some sort of `universal SCFT,' a physical theory 
in which superconformal deformation parameters have been promoted to
non-dynamical fields covering the entire moduli space or stack, 
such that one gets a unique ordinary SCFT
at any point of the moduli stack by `freezing' the superconformal deformation
parameters to an appropriate value.
In QFT terms, we do not know a precise definition of such a structure.
However,
in mathematics, such structures are often studied, and constraints
on the existence of such an object are well-known.  

In this section,
we will apply some of the general principles known in mathematics
to give a brief overview of possible obstructions to the existence
of such a hypothetical universal SCFT structure, and associated issues.

In general, in constructing a moduli space of objects of some type,
it sometimes happens that to a single point on the moduli space can be
associated multiple objects.  For example, in moduli spaces of
holomorphic vector bundles, there can be loci along which several
semistable bundles are associated to the same point on the moduli space
(see {\it e.g.} \cite{es-kcs,aglo1,aglo2} and references therein).
This issue arises in heterotic strings; however, of the semistable bundles
over any one point, only one defines a SCFT, the others define massive
theories.  In principle, analogues also exist at singular limiting
points in moduli spaces of complex structures, see {\it e.g.}
\cite{fm}.

If a single point on the moduli space can be associated to multiple objects,
then, it is not possible to construct a universal object over the moduli space.
However, intuitively this only happens at singular points on the moduli
space, so if we restrict to smooth points, it seems plausible that
we can avoid this issue.

In principle, another issue that can arise involves objects with
additional automorphisms.  If one walks around the moduli space and
returns to oneself but only up to an automorphism, it may not be
possible to find a universal object.  One way to resolve this issue is to
change the moduli problem by adding extra data that removes the automorphisms,
hence `rigidifying' the problem.  A different approach that does not involve
changing the objects themselves is to
replace moduli spaces with moduli stacks, in which the automorphisms
are essentially divided out.  We have speculated about such stacks in the
case of moduli spaces of elliptic curves in section~\ref{sect:ex:t2};
however, we should also emphasize that, due to the fact that SCFTs can have
in principle many subtle symmetries, including higher group
symmetries (see {\it e.g.} \cite{genlglob}), it would be difficult to completely
settle this issue for moduli spaces of SCFTs at this time.

One of the subtleties we see in the example of moduli spaces of
elliptic curves in section~\ref{sect:ex:t2}
is that even if a universal SCFT exists, it may not
be possible to construct operators that extend globally across the
moduli space.  For example, as discussed in section~\ref{sect:ex:t2},
the spectral flow operator ${\cal U}_{1/2}$ is locally a section of the
Bagger-Witten line bundle over a gerbe over the moduli space of 
elliptic curves.  However, the Bagger-Witten line bundle is not an honest
line bundle, it is only a fractional line bundle, and as such, admits no
meromorphic sections, neither over the stack nor over the underlying coarse
moduli space.  Even if we omit the points with enhanced stabilizers,
the restriction of the Bagger-Witten line bundle to the complement is
still fractional and still does not admit meromorphic sections,
short of performing some noncanonical transformation to remove the
twisting.  As a result, because there is no global meromorphic
section, it is not possible to unambiguously associate
an operator ${\cal U}_{1/2}$ to every point of the moduli space -- we can
make local choices, but there does not exist a global consistent choice.

In passing, it is tempting to apply ideas about existence of universal
structures to F theory compactifications, to argue for the necessity
of non-mutually-local branes.  Consider an F theory compactification
to six dimensions, on $T^2 \times K3$, along the fibers of
an elliptically-fibered $K3$.  
Morally the ${\mathbb P}^1$ base of the $K3$ is parameterizing a family
of elliptic curves, which (because we have compactified down to six dimensions),
we can think of has a family of $N=2$ SCFTs.  If a universal SCFT
structure existed over that entire parameter space, that fact might
suggest that the fibration should be trivial, admitting no monodromies.
However, the existence of non-mutually-local branes means that we do not
have a global family of SCFTs, and so such an argument does not apply.
In general terms, this appears to be broadly consistent 
with observations about monodromies resulting from
non-mutually-local branes in \cite{ferrara}[section 6].

Given that building a well-behaved
moduli space of elliptic curves requires
working with stacks instead of spaces, one might wonder whether 
stacks suffice when describing universal objects over moduli stacks of
SCFTs.  One possibility is that one might have to work with higher stacks,
a suggestion which is implicit in the proposed construction we outline
next.

\subsection{Sheaves of RG flows}

In passing, let us very briefly outline a proposal for another approach to
constructing possible `universal SCFT's.'

Cover the CFT moduli space by open patches $\{ U_{\alpha} \}$.
To each patch $U_{\alpha}$, 
associate a family of quantum field theories (not 
necessarily conformal field theories), in which deformation parameters
have been promoted to non-dynamical fields, and which in principle should
RG flow to conformal field theories associated with points on 
$U_{\alpha}$.
To `glue' these together on overlaps $U_{\alpha} \cap U_{\beta}$,
we require that the two quantum field theories RG flow to the same
theory at some scale $\Lambda_{\alpha \beta}$, modulo field redefinitions
and irrelevant operators.  On triple overlaps $U_{\alpha} \cap U_{\beta}
\cap U_{\gamma}$, there is no guarantee
that the three scales $\Lambda_{\alpha \beta}$, $\Lambda_{\beta \gamma}$,
$\Lambda_{\gamma \alpha}$ or their associated field redefinitions or
irrelevant operator insertions will all be comparable, so we must instead
require that on triple overlaps, the three quantum field theories 
associated to the three patches will all RG flow to the same theory
at some scale $\Lambda_{\alpha \beta \gamma}$, modulo field redefinitions
and irrelevant operators.  One would of necessity continue to impose the
same constraint on quadruple and all higher overlaps.

In broad brushstrokes, by associating multiple different quantum field
theories to each open set, one would get a type of infinity sheaf in
this fashion,
with 
gluing conditions provided by renormalization group flow (and field
redefinitions and irrelevant operators) on intersections.  A potential
`universal SCFT' would be a global section of that sheaf.

In this language, the idea that the spectral flow operator ${\cal U}_{1/2}$
is a section of a fractional line bundle, is tied to the idea that under
renormalization group flow, a UV $R$ symmetry may mix with non-R symmetries,
as has been applied recently in {\it e.g.} $c$-extremization \cite{bb}.
Intuitively, the IR $R$ symmetries in different patches on the moduli space
might subtly differ if there are other symmetries in the theory, and in 
particular could fail to close on triple overlaps.

\section{Left-moving analogues in heterotic compactifications}
\label{sect:left-bundle}

So far in this paper we have primarily focused on Bagger-Witten
structures in (2,2) supersymmetric worldsheet SCFTs.
In this section we will briefly outline how structures analogous to the
Bagger-Witten line bundle can arise on moduli spaces of (0,2) SCFTs.

In a typical perturbative heterotic compactification on a Calabi-Yau,
involving an $SU(n)$ gauge bundle,
the left-moving fermions also admit a $U(1)$ symmetry (which becomes\footnote{
Just as ordinary mirror symmetry flips the sign of charges under the
left-moving $U(1)_R$, (0,2) mirror symmetry is similarly defined 
by a sign flip in charges under
this canonical $U(1)$.
} the
left-moving $U(1)_R$ on the (2,2) locus), which rotates the left-moving
fermions by phases.  One might ask about the analogue
of Bagger-Witten for this symmetry.  In fact, that left-moving $U(1)$ forms
part of the low-energy gauge symmetry.  For example, in a compactification
of an $E_8 \times E_8$ string on a Calabi-Yau threefold with an $SU(3)$
bundle, the first $E_8$ is broken to an $E_6$, which is realized on the
worldsheet via a Spin$(10)\times U(1)$ subalgebra.  The $U(1)$ factor is
precisely the left-moving $U(1)$ described above.  In any event, as one
walks around the SCFT moduli space, the left-movers come back to themselves
up to Spin$(10) \times U(1)$ and, ultimately, $E_6$ rotations.  Thus,
if one has low-energy gauge symmetry $G$, then over the SCFT moduli space,
target-space fields behave as sections of a vector bundle associated to
a principal $G$ bundle, of which the $U(1)$ above defines a rank one subbundle.

Part of the reason for our focus on (2,2) theories is that, as
explained previously in section~\ref{sect:constr-Liouville},
moduli spaces of (0,2) SCFTs have multicritical points at which
varieties of different dimension intersect one another, and so
are not amenable to analyses which assume the moduli space is smooth.
Our observations above, however, do not require the moduli space to 
be smooth.

\section{Conclusions}

In this paper we have examined recent claims about the Bagger-Witten
line bundle in the context of a specific concrete example, namely the
moduli space of elliptic curves.  We have shown how in general the
Bagger-Witten line bundle is not actually an honest line bundle,
but rather will only be `fractional,' as in fact happens over 
moduli spaces of elliptic curves.  As a fractional line bundle,
it is nontrivial but torsion, which is consistent with recent results
of \cite{gkhsst} stipulating that consistency of the worldsheet
metric in families of SCFTs requires that the Bagger-Witten line bundle
admit a flat connection.  We have also examined more general features of such
`universal' constructions of families of SCFTs in two dimensions,
and proposed another construction utilizing sheaves of
massive quantum field
theories.
Finally, we have discussed other bundles appearing over moduli spaces
of SCFTs.

\section{Acknowledgements}

We would like to thank L.~Anderson, P.~Aspinwall, P.~Candelas, 
J.~Distler, R.~Donagi, X.~Gao,
J.~Gray, R.~Hain, S.~Katz, Z.~Komargodski, S.~J.~Lee, I.~Melnikov,
D.~Morrison, R.~Plesser,
M.~Rocek, and S.~Theisen
for useful conversations, and especially T.~Pantev for many discussions
of stacks and derived algebraic geometry.
E.S. was partially supported by NSF grant PHY-1417410.

\end{document}